\documentclass[11pt,preprint,flushrt]{aastex}
\usepackage{amsmath}
\usepackage{amssymb}
\usepackage{latexsym}
\usepackage{cancel}
\usepackage{verbatim}
\usepackage{indentfirst}
\usepackage{multirow}
\usepackage{color}
\usepackage{graphicx}
\usepackage{multicol}
\usepackage[colorlinks=true, allcolors=blue]{hyperref}

\usepackage[english]{babel}
\usepackage[utf8x]{inputenc}
\usepackage[T1]{fontenc}

\usepackage[a4paper,top=3cm,bottom=2cm,left=3cm,right=3cm,marginparwidth=1.75cm]{geometry}

\usepackage[colorinlistoftodos]{todonotes}

\begin{document}

\newcommand {\gs} {{\tt{GalSim}}}  
\newcommand {\wf} {WFIRST}  
\newcommand {\bfe} {BF}

\newcommand{\red}[1]{\textcolor{red}{#1}}
\newcommand{\blue}[1]{\textcolor{blue}{#1}}
\newcommand{\brown}[1]{\textcolor{brown}{#1}}
\newcommand{\orange}[1]{\textcolor{orange}{#1}}

\title{Laboratory measurement of the brighter-fatter effect in an H2RG infrared detector}
\author{A. A. Plazas$^{\dagger a}$, C. Shapiro$^a$, R. Smith$^b$, E. Huff$^a$, \& J. Rhodes$^{a,c,d}$}
\email{$^\dagger$plazas@caltech.edu}
\affil{$^a$Jet Propulsion Laboratory, California Institute of Technology, 4800 Oak Grove Dr., Pasadena, CA 91109, USA}
\affil{$^b$Caltech Optical Observatories, 1200 E. California Blvd, CA 91125, USA}
\affil{$^c$California Institute of Technology,\\1200 E. California Blvd., CA 91125, USA}
\affil{$^d$Institute for the Physics and Mathematics of the Universe, 5-1-5 Kashiwanoha, Kashiwa, Chiba Prefecture 277-8583, Japan}
\begin{abstract}
{The “brighter-fatter” (BF) effect is a phenomenon---originally discovered in charge coupled devices---in which the size of the detector point spread function (PSF) increases with brightness.  We present, for the first time, laboratory measurements demonstrating the existence of the effect in a 
Hawaii-2RG HgCdTe near-infrared (NIR) detector.  We use JPL's Precision Projector Laboratory, a facility for emulating astronomical observations with UV/VIS/NIR detectors, to project about 17,000 point sources onto the detector to stimulate the effect.  After calibrating the detector for nonlinearity with flat-fields, we find evidence that charge is nonlinearly shifted from bright pixels to neighboring pixels during exposures of point sources, consistent with the existence of a BF-type effect.  NASA’s Wide Field Infrared Survey Telescope (WFIRST) will use similar detectors to measure weak gravitational lensing from the shapes of hundreds of million of galaxies in the NIR.  The WFIRST PSF size must be calibrated to $\approx$ 0.1\% to avoid biased inferences of dark matter and dark energy parameters; therefore further study and calibration of the BF effect in realistic images will be crucial.}
\end{abstract}

\section{Introduction}
Weak gravitational lensing (WL) of the large-scale matter distribution of the Universe has been identified as one of the main techniques to probe the nature of dark matter and dark energy \citep{albrecht06,hoekstra08,kilbinger15}. Images of distant sources are subtly distorted and magnified by the dark matter distribution along the line of sight. This ``cosmic shear'' signal has been measured by past and current ground-based galaxy surveys to constrain the current standard cosmological model \citep{desy1all,troxel17,kuijken15}. In the future, projects such as the Large Synoptic Survey Telescope (LSST) \citep{ivezic08}, ESA's Euclid spacecraft \citep{laureijs11}, and NASA's Wide Field Infrared Telescope Survey (WFIRST) space mission \citep{spergel15} will measure the shapes of hundreds of millions of galaxies to use WL as a cosmic probe. 

However, the signal induced by cosmic shear is small (1--2\% distortion) and must be measured to a fraction of percent to produce cosmological constraints with the precision required by galaxy surveys. As a consequence, great effort has been placed to understand and characterize the systematic errors that arise during the galaxy shape measurement process. These errors may arise from different sources, e.g., the atmosphere, the telescope optics, galaxy shape measurements, correction for the Point Spread Function (PSF) of the system, redshift distributions, intrinsic alignments of galaxy shapes due to tidal fields, instrumental signatures on the data, etc. \citep{mandelbaum17,zuntz17,jarvis14,mandelbaum14b,stubbs14} 

{WL surveys typically use Charged Coupled Devices (CCDs) in their focal planes.} One of the subtle systematic errors that originate in {CCDs} is known as the ``brighter-fatter'' effect (BF). This is a nonlinear effect initially measured as a deviation from the linearity in the photon transfer curve of flat field exposures (signal variance as a function of mean signal level), indicating a departure from Poissonian statistics \citep{downing06}. The effect also manifests as an increase in the size of point sources---such as stars--- with flux (hence the name), due to the deflection of new charges induced by the electric field generated by previously accumulated charge in a given pixel. 

The BF effect has been observed {in the CCD's of} several wide-field cameras such as the Dark Energy Camera (DECam, \citet{flaugher15,gruen15}), Megacam \citep{guyonnet15}, the Hyper Suprime-Cam \citep{mandelbaum17b}, and in the prototype sensors of future projects such as LSST and Euclid \citep{baumer15,lage17,niemi15}. It has been measured to be linear in flux, slightly asymmetric in the channel and serial register directions, achromatic, and long range (e.g., up to 10 pixels in DECam, \citet{gruen15}). The BF effect breaks the assumption in astronomical image analysis that the apparent sizes of stars are independent of their brightnesses. Since bright stars are used to determine the PSF, the measured PSF will be systematically larger than the correct PSF that is convolved with the faint galaxies used for WL analyses, {inducing shear calibration errors. \citet{mandelbaum14b} calculate that a 1\% mis-estimation in the PSF size caused by the BF effect will induce a multiplicative shear error of $\sim \text{0.02--0.06}$. Likewise, a 1\% ellipticity error will result in an additive term of about $3\times 10^{-4}$. In addition, \citet{huterer06} show that---for a spaced-based mission---a multiplicative error of $\sim 0.04$ will result in a 60\% degradation in the marginalized error of the dark energy equation-of-state parameter $w_0$. Consequently, Stage IV dark energy surveys (as defined by \cite{albrecht06}) demand a knowledge of the relative PSF size and ellipticity better than 10$^{-3}$ \citep{amara08,paulin08,paulin09,cropper13,massey13,spergel13}.}

One way to mitigate the BF effect is to exclude the brightest stars from the shape measurement pipeline (e.g., \citet{jarvis15,zuntz17}); however, this is not an ideal solution. \citet{antilogus14} have developed a phenomenological model in which the boundaries\footnote{The pixel boundaries in an astronomical detector do not refer to a physical barrier, but are determined by electric fields that define a plane that indicates by which pixel a charge will be captured.} of a particular pixel shift by an amount proportional to the charge already stored in that pixel and its neighbors. The parameters of the model can be derived from measuring correlations in flat-field images and by assuming certain symmetries in the problem. The model has been shown to be consistent with data from different types of CCDs \citep{guyonnet15}, and has been implemented by \citet{gruen15} in the publicly available {\tt{GalSim}} code \citep{rowe15}. In the case of the Hyper Suprime-Cam survey, \citet{mandelbaum17b} apply a correction that assumes that the electric field transverse to the main drift field can be expressed as the gradient of a potential proportional to the accumulated charge, and demonstrate that the BF-effect is attenuated below scientific requirements \citep{coulton17}.   

The future space mission WFIRST will use near-infrared (NIR) detectors, which have {a different architecture} compared to CCDs and which have not yet been used to perform galaxy shape measurements at the required accuracy for WL science. In particular, the Wide Field Imager of WFIRST will use an array of 18 H4RG-10\footnote{HAWAII---HgCdTe Astronomical Wide Area Infrared Imager---Reference pixels and Guide mode.} (with a pixel pitch of 10 $\mu$m and a size of 4096 pixels on each side) hybrid complementary metal-oxid-semiconductor (CMOS)  devices manufactured by Teledyne \citep{beletic08}. Unlike the semiconductor material used in CCDs, the light detecting part of these detectors consists of HgCdTe (mercury cadmium telluride). Photo-generated charges are collected in the depletion region generated at the p-n junction at the detector layer, inducing a change in voltage that is read through non-destructive Fowler sampling or ``up-the-ramp" sampling.  The detector layer is connected to a readout integration circuit (ROIC or multiplexor) layer, which transmits the charge to off-chip electronics for digitization. 

While the BF effect is well-known in CCDs, it has not been studied in detail in NIR detectors. Despite their physical differences with CCDs, \citet{plazas17} (P17 hereafter) propose that there are physical motivations to expect a BF effect in CMOS NIR detectors. As a pixel accumulates charge, the detector substrate voltage changes, and, as a consequence, the pixel's local depletion region shrinks.  If it shrinks significantly relative to a neighboring pixel, e.g. due to a local concentration of flux in an image, then new charge generated close to the midpoint between the two pixel centers may have a higher probability of being collected in the pixel with the larger depletion region.\footnote{Note that since the initial depletion width is set by the detector substrate voltage, and since quantum efficiency and inter-pixel capacitance (IPC) are insensitive to that voltage, it is reasonable to assume that changes in the depletion width as signal accumulates will not affect total quantum efficiency or IPC.}
Thus, the pixels' collection zones physically shift depending on the signal contrast between adjacent pixels.  One consequence is that point source images would experience a fluence-dependent redistribution of charge away from the brightest central pixels, i.e. brighter sources appear fatter.  This proposed mechanism is in contrast to the model for CCDs, in which incoming charge is deflected away from pixels with high charge concentrations due to changes in the direction of the local electric field (in this model, the effect can range across several pixels).  
Early indications of a BF effect were seen in archival images of the globular cluster Omega centauri ($\Omega$-cen), obtained with the infrared H1RG-18 device of the Wide Field Camera 3 (WFC3-IR) on the Hubble Space Telescope. By analyzing the central 3$\times$3 region of bright (but unsaturated) stars, P17 find evidence for charge redistribution from the central pixel to neighboring pixels as the integration time increases.\footnote{P17 follows an analysis by Jay Anderson (STScI), who presented preliminary results on the BF effect with  the same type of data ($\Omega$-cen images taken with the WFC3-IR channel) at a WFIRST meeting in May of 2016.} 

In this work we use data from Precision Projector Laboratory (PPL) \citep{shapiro18,seshadri13,shapiro13} to study the BF effect in NIR detectors, extending the initial analysis performed in P17.
The PPL is a facility run by NASA’s Jet Propulsion Laboratory (operated by the California Institute of Technology) with support from Caltech Optical Observatories (COO).  It was designed to assess the impact of detector effects on space-based astronomical observations such as weak lensing shape measurements. {Laboratory measurements provide several advantages over on-sky data, which suffer from effects such as drift, thermal instability, and source confusion. Tests at the PPL can be repeated numerous times under controlled conditions, allowing us to achieve a more reliable measurement. The PPL test-bed (``the projector'') is stable and versatile, with a wide range of control of the flux, wavelength, position, and f-number of the projected images. Images (e.g., a grid of point sources) can be focused with high Strehl ratio over the entire area of a detector, which provides a multiplexed advantage over e.g. scanning a single bright spot. }

This paper is organized as follows: in Section 2 we describe the laboratory setup and parameters, as well as the characteristics of the NIR detector used to acquire the data analyzed. In Section 3 we summarize the PPL data calibration and analysis steps, including correction for known effects such as nonlinearity in the conversion from charge to voltage (known as ``voltage nonlinearity" or ``classical nonlinearity"---NL;\citet{hilbert14,hilbert08,plazas16}) and electronic coupling between pixels (known as ``inter pixel capacitance"---IPC; \citet{hilbert11,mccullough08}). We also present null tests that validate our measurement pipeline using simulations. Results are shown in Section 4. Finally, we conclude in Section 5. 
\begin{figure}[htbp]
\centering
  \begin{minipage}[b]{0.4\textwidth}
    \includegraphics[width=0.8\textwidth]{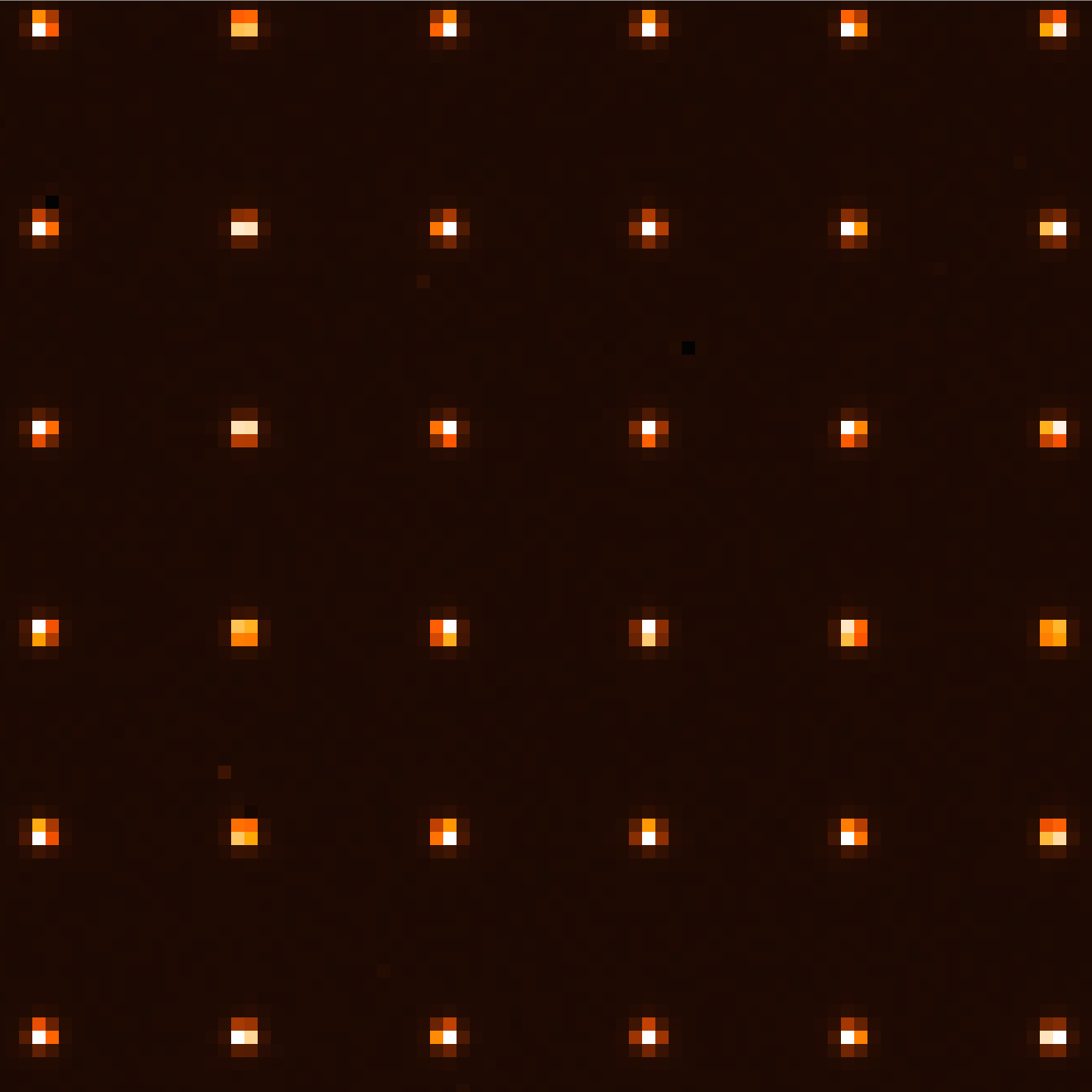}
  \end{minipage}
  \begin{minipage}[b]{0.35\textwidth}
    \includegraphics[width=0.923\textwidth]{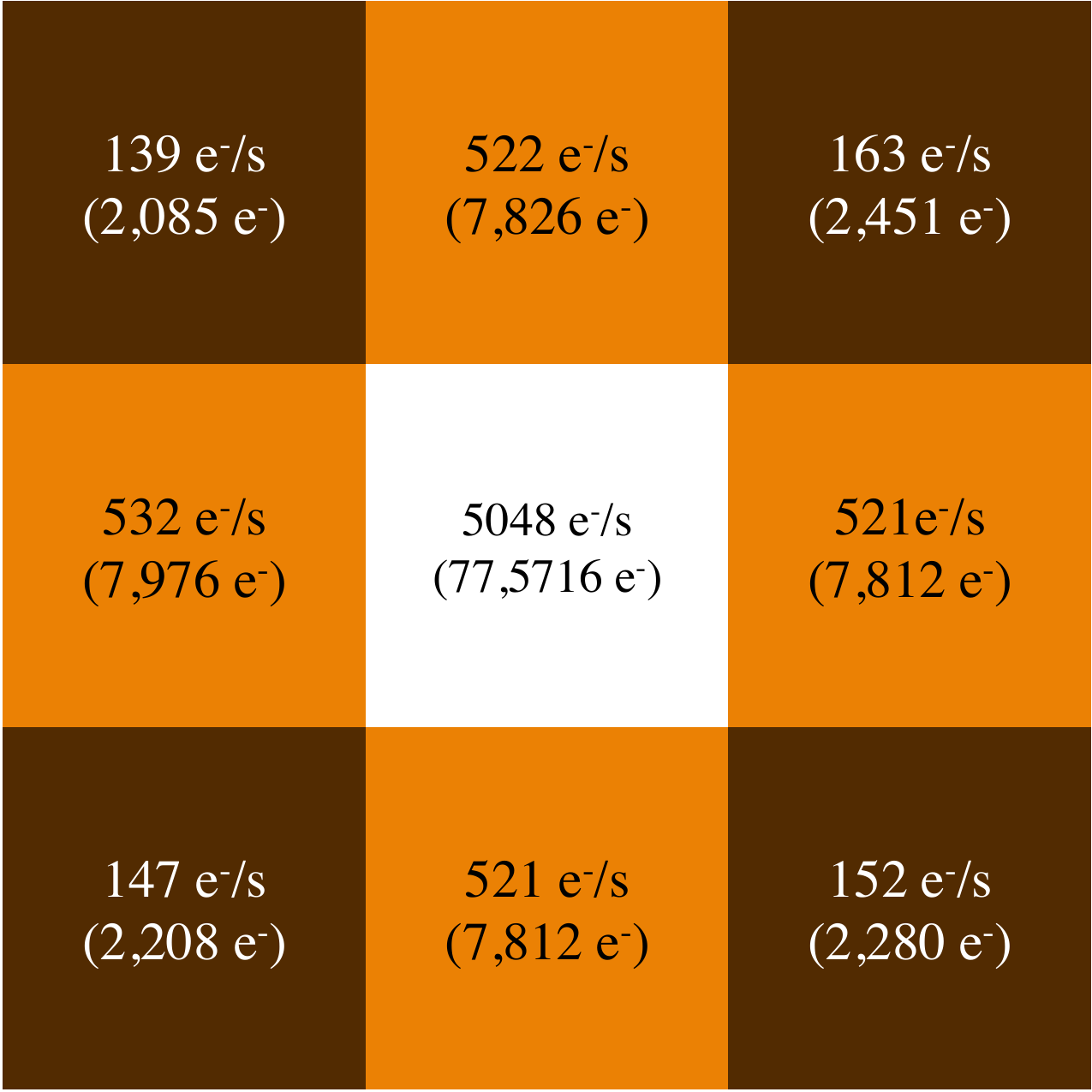}
  \end{minipage}
\caption{\label{fig:spots} \emph{Left}: Subsection of a raw image showing a grid of $\sim$ 17000 point sources projected on a 2k$\times$2k H2RG detector at the PPL. \emph{Right}: Postage stamp showing the inner 3$\times$3 pixels region of the PPL spots, obtained by taking the median over approximately 700 sources with a centroid $< 0.1$ pixels away from the pixel center. Each point in the grid is strongly under-sampled (FWHM < 1 pixel).}
\label{fig1}
\end{figure}

\section{Data and simulations}

\subsection{Precision Projector Laboratory Data}


The detector tested in this experiment was an engineering-grade Teledyne H2RG-18 NIR detector ($2048^2$ pixels; 18 $\mu$m pitch) with a cutoff wavelength of 2.3 $\mu$m.  The NIR detector working group for the Euclid mission lent the H2RG to the PPL for the purpose of investigating intra-pixel response variations, which impact the photometric accuracy of the Near Infrared Spectrophotometer (NISP; \citet{maciaszek16,prieto12}).  That study will be presented in a separate paper.  The H2RG (SN 18546) is designated engineering-grade primarily due to a large collection of inoperable pixels in one corner.  It is otherwise an unexceptional device, and we presume that the BF effect results presented here will be qualitatively representative of other HxRG devices, in particular the H4RG-10 detectors planned for the WFIRST imaging survey.  Conventional performance characterization of the H2RG was conducted by Teledyne before transfer of the device to the PPL.

We use the PPL test-bed to project a grid of approximately 17000 point sources (``stars'' or ``spots'') on the H2RG with a uniform spacing of 274.5 $\mu$m (15.25 pixels), illuminated by a broadband quartz-tungsten-halogen lamp with Y-band filter (0.9--1.07 $\mu$m). Setting the focal ratio to f/11 produces a diffraction limited PSF with a full-width-half-maximum (FWHM) of about 11 $\mu$m, broadened to about 14 $\mu$m (0.78 pixels) by lateral charge diffusion and seeing.  The PSF flux is thus strongly concentrated within the area of a single pixel. Images are generated by illuminating a $56$ mm $\times$ $56$ mm $\times$ $2.3$ mm quartz target mask with pinholes etched into a chrome coating (manufactured by HTA Photomask Inc.). The pinholes have a diameter of approximately $3\ \mu$m, which is unresolved by the 14 µm FWHM PSF and thus insensitive to small fabrication errors.  Note that although we rely on highly concentrated point sources to stimulate the BF effect, we do not use knowledge of the spot profiles or the PSF in this particular analysis. Fig. \ref{fig:spots} shows an image of the spot grid on a subregion of the detector and a typical (median) flux distribution of the central 3$\times$3 pixels when a spot is centered on a pixel.

The detector was operated at a temperature of $95$ K with a room-temperature Leach controller clocked at 166 kHz.  In 32-channel readout mode, the maximum frame-rate is about 1.2 Hz for the full detector area, but in practice we sampled at 0.33 Hz to avoid a data writing glitch (which has since been fixed).  Fluxes are measured by "sampling up the ramp", i.e. non-destructively reading each pixel at regular (3 s) intervals during an exposure. During non-destructive reading, the photo-generated charges are collected on the diode capacitance, causing a drop in the initial reverse bias of the pixel. The voltage in the output metal oxide semiconductor field effect transistors (MOSFET) is read as the charges accumulate, and any change in voltage with time can be measured (sampled) without affecting or destroying the charge on the diode \citep{rieke07}. Thus, after applying a voltage-to-charge calibration to an exposure, we obtain a time-series of accumulated charge for each pixel, as opposed to e.g. a CCD which only provides each pixel's final accumulated charge. 

For each ramp, we discard the first frame sampled immediately after resetting to mitigate transient effects caused by the reset.  Thus with 6 frames per ramp and a total exposure time of 15 seconds, we analyze the final 5 frames (12 s). We apply a mean conversion gain of 2.7 e$^-$/ADU and measure a median read noise of 15 e$^-$ (correlated-double-sampling noise for 2 consecutive frames).

For calibration purposes, dark frames (which include thermal background and light leakage) and flat-field images (uniform illumination) are taken with the same frame-rate and exposure time as the spot images. Our data set consists of 100 spots and flat-field ramps, and 10 dark-frames ramps.  Under the illumination conditions, the average central pixel of a spot reaches $\sim$ 76000 e$^-$ in an exposure (about 5000 e$^-$/s) while for the flat-field images it reaches $\sim$ 95500 e$^-$(about 6400 e$^-$/s) after correcting for NL. These correspond to 57\% (spots) and 72\% (flats) of the 132245 e$^-$ mean well depth (defined by onset of 4\% NL; measured by Teledyne).  With a mean background level of  {94 e$^-$/s/pixel}, the exposures are shot noise dominated with a signal-to noise ratio (SNR) of about 120 per frame for the central pixel of a spot.


\subsection{Simulations}
We also simulate exposures that resemble the lab data in order to asses the performance of our measurement pipeline and characterize its response to different parameters. We use the code {\tt{GalSim}}\footnote{\url{https://github.com/GalSim-developers/GalSim}} (v. 1.3.0) to create 90 spots ramps with a grid of 1024 point sources on a 2k by 2k pixels image, using an empirical PSF model that matches the diffraction-limited PSF of our experimental setup. We use this grid of spots as input for a simulator that produces ramps that have the same number of frames and integration time of the laboratory data (6 and 15 seconds, respectively). We also generate 90 flat and dark ramps. The simulator includes the option of applying different sources of noise such as shot noise and readout noise (set to 15 e$^-$ r.m.s.), IPC, classical nonlinearity (NL), and the BF effect model \citep{antilogus14,gruen15} built-in in {\tt{GalSim}}.\footnote{{\tt{galsim.cdmodel}}} This model is phenomenological and independent of the underlying physics of the effect. We invoke this model in order to generate simulated spots that empirically approximate our data for the purpose of estimating noise. 


\section{Methods}

As in P17, we record the inner $3\times3$ pixel region (a ``postage stamp") of each spot in order to look for evidence of the BF effect in the PPL data. An advantage of non-destructive reads is that we can compare the fluxes in each spot at the end of an exposure to flux at the beginning, instead of comparing different spots (which adds PSF noise) or different exposures (which adds shot noise). We look for deviations from constant flux after correcting for known effects such as NL and IPC. We calibrate the NL per pixel by using flat-field images, where we expect the contribution to nonlinearity in the signal due to the BF effect to be suppressed, given that the signal contrast between pixels is small for a mean flat-field image. After correcting NL in the spot images, we look for evidence of non-constant flux due to the charge contrast between pixels.

We discard the first 10 ramps of the initial set of 100 spot and flat-field ramps to mitigate the impact of the ``burn-in'' effect, an effect---different from persistence \citep{smith08,biesiadzinski11}---in which the sensitivity of the detector to light is increased following previous exposures.\footnote{This effect has been observed in H4RG-10 data taken at the Detector Characterization Laboratory in Goddard's Space Flight Center.} In addition, we have visually inspected the ramps formed by taking the mean of a region of size 1800$\times$1800 pixels in each frame for the dark, flat, and spot exposure types, and have discarded additional outliers ramps. The remaining number of ramps is thus 9, 85, and 84 for the darks, flats, and spots, respectively. 

We then calculate the median ramp of each set in order to increase the SNR. In addition to improvements in SNR, stacking the ramps averages over lamp fluctuations\footnote{Previous photometry experiments at the PPL have demonstrated stability in the brightness of point sources to the level of 1 part in 10$^4$ over observing series of hours \citep{touli14}.} and other systematic errors such as image motion and seeing ({\citet{shapiro18} show that the mean position across the detector has a displacement of about 1 $\mu$m r.m.s. for a single image.}). We run {\tt{SExtractor}} \citep{bertin06} on the sample with the longest integration time of the median spots ramps, using a mask flagging bad pixels produced with laboratory measurements. In this way, we identify the approximate location of the spots in our images, and calculate the unweighted centroid from the inner 3$\times$3 postage stamp of each spot. 
\begin{figure}[htbp]
\centering 
\includegraphics[width=1.0\textwidth, page=5]{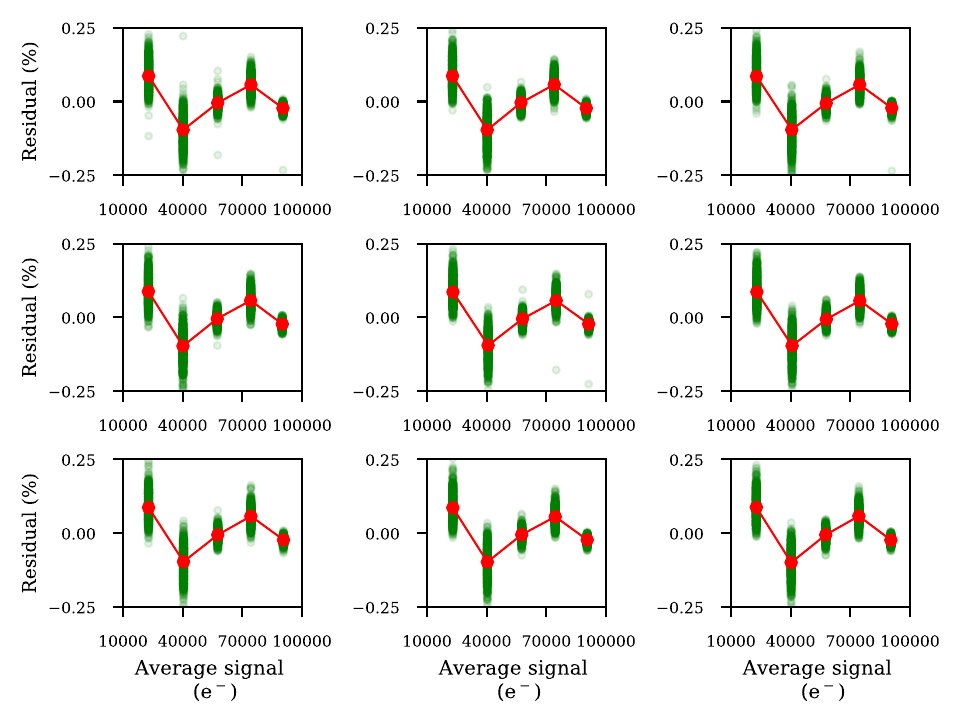}
\caption{\label{fig:res} Fractional NL residuals in flat-field images. The pixel positions of the center of the spots are used in the stacked flat-field image to select 748 3$\times$3 pixels regions (a postage stamp) around that center. Then, the signal---before dark subtraction---in each of the 9 pixels of each region is fitted to Eqn. \ref{eq:NL}, and the relative residuals per frame are reported in the y-axis (green spots). The  x-axis shows the mean signal averaged over the 9 pixels of each region. The red line represents the median value per frame.}
\end{figure}
We initially select those sources with a centroid within $\pm$ 0.1 pixels of the pixel center to maximize the signal contrast (and thereby increasing the BF amplitude) between the central pixel of each spot and its nearest neighbors. 

We correct for IPC in order to improve our estimates of charge contrast in neighboring pixels.  We assume that IPC is linear as well as spatially and temporally uniform, and we correct for it by applying the following kernel, estimated by comparing the signal in hot pixels to that of their nearest neighbors \citep{hilbert11,mccullough08}: 
\begin{equation}
K=
\begin{pmatrix}
0 & -0.007 & 0 \\ 
-0.009 & 1.032 & -0.009 \\
0 & -0.007 & 0 \\
\end{pmatrix}
\label{eqn:k}
\end{equation}
An exact correction is not crucial here since linear IPC is easily distinguishable from the BF-effect---the former redistributes a fixed fraction of a pixel signal to its neighbors and would not change with exposure time or contrast.  Nonlinear IPC, however, could induce its own ``brighter-fatter'' effect and either mimic (or cancel out, depending on the sign) the nonlinear effects of actual charge shifting.  A detection of either effect on the PSF is interesting and important to characterize for precision astronomical measurements, and we do not attempt to separate them in this initial investigation.  Donlon et al. \citep{donlon16,donlon17} have observed effects around hot pixels (with signals below $\sim$ 15,000 e$^-$) in H2RGs which they ascribe to nonlinear IPC. Complementary measurements of IPC nonlinearity are underway by the WFIRST detector working group using isolated pixel resets \citep{seshadri08} and noise correlations between adjacent pixels.

We convert the ADU numbers from raw images to electrons by using an IPC-corrected average mean of 2.7 e$^-$/ADU. We then fit a quadratic polynomial to the signal ramp in each pixel, S(e$^-$), in order to correct for classical voltage nonlinearity:
\begin{equation}
S= C_0 + C_1t + C_2(C_1t)^2
\label{eq:NL}
\end{equation}
In Eq. \ref{eq:NL}, $C_0$ represents an offset in the ramp due to a reset voltage, and $C_1t \equiv Q_{\mathrm{L}}(t)$ is the linear flux component of the signal. P17 show (their Appendix A) that the dependence of capacitance on voltage in the pixel diode leads to an approximate quadratic response at low to medium signals. Figure \ref{fig:res} shows that this quadratic polynomial model correctly describes flat-fielded data to $0.1\%$ (on average), up to a signal level of approximately $95000$ e$^-$ ({$72\%$} of the mean full-well of the pixels of the detector). 

In terms of the parameters of the fit, the corrected signal $Q_{\mathrm{L}}(t)$ (the signal that would have been detected by a linear detector) is given by solving the quadratic formula as in \citet{kubik14}:
\begin{equation}
Q_{\mathrm{L}}(t) = \frac{1}{2C_2}(-1 + \sqrt{1-4C_2(C_0 - S)})
\label{eq:correction}
\end{equation}
We use the quadratic coefficient $C_2$ fitted from the flat-field images {to correct the spot and dark ramps for NL}. After subtracting the corrected dark signal from the corrected spot signal for each pixel in the $3\times3$ postage stamp, we convert the signal in each pixel ramp (in e$^{-}$) into a flux rate (in 1/e$^{-}$ per seconds). 
{Finally, to quantify post-calibration nonlinearity when switching from flat-field images to spot images---an indicator of the BF effect---we define the relative flux deviation in a pixel as follows:
\begin{equation}
f_{\mathrm{N}}(k) = \frac{F_k - F_1}{F_*}
\label{eq:fn}
\end{equation}
where $F_{k}$ is the pixel flux calculated by differencing the $k$th and $(k+1)$th frames (the discarded reset frame is $k=0$), and $F_*$ is the time-averaged total flux of the 9 central pixels of the spot.  Eqn. \ref{eq:fn} differs from the metric used in P17 by the normalization $F_*$, which in this case is defined as:
\begin{equation}
F_*=\frac{\sum_{p} ( S_{p,M-1} - S_{p,1} ) }{ \Delta t(M-2)}
\label{norm}
\end{equation}
where $S_{p,k}$ is the signal in the $k$th frame of the $p$th pixel, $M=6$ is the initial total number of frames, and $\Delta t = 3$ s is the frame time.  The sum is over the 9 central pixels.}


\section{Results}
\subsection{Measurement of flux deviations in point source images}
\begin{figure}[htbp]
\centering 
\includegraphics[width=1.0\textwidth, page=2]{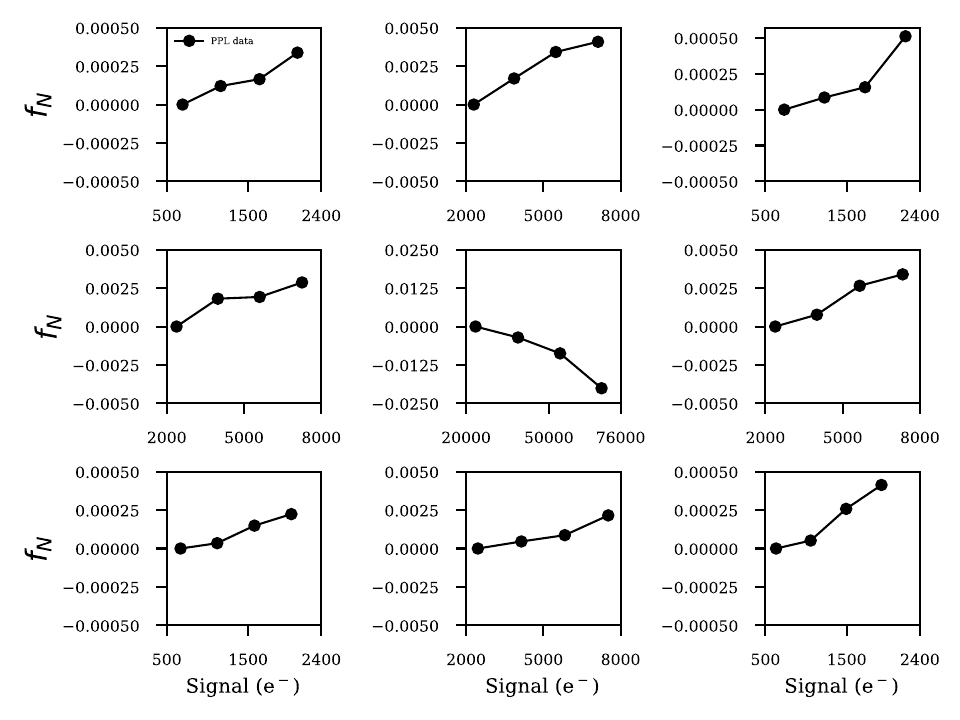}
\caption{\label{fig:9pixels} For each pixel in a 3$\times$3 region containing a spot, we plot the average $f_N$ -- the change in flux relative to the start of the exposure, normalized by the average total flux of the spot.  The data is averaged over 84 exposures and over 748 spots with a centroid $ <0.1$ pixels away from a pixel center. $f_N$ is plotted as a function of the mean integrated signal between the two consecutive frames from which flux is estimated. The error bars, which are computed from the standard error of the mean taken over all spots, are smaller than the symbols.}
\end{figure}

\begin{figure}[htbp]
\centering 
\includegraphics[width=1.0\textwidth, page=7]{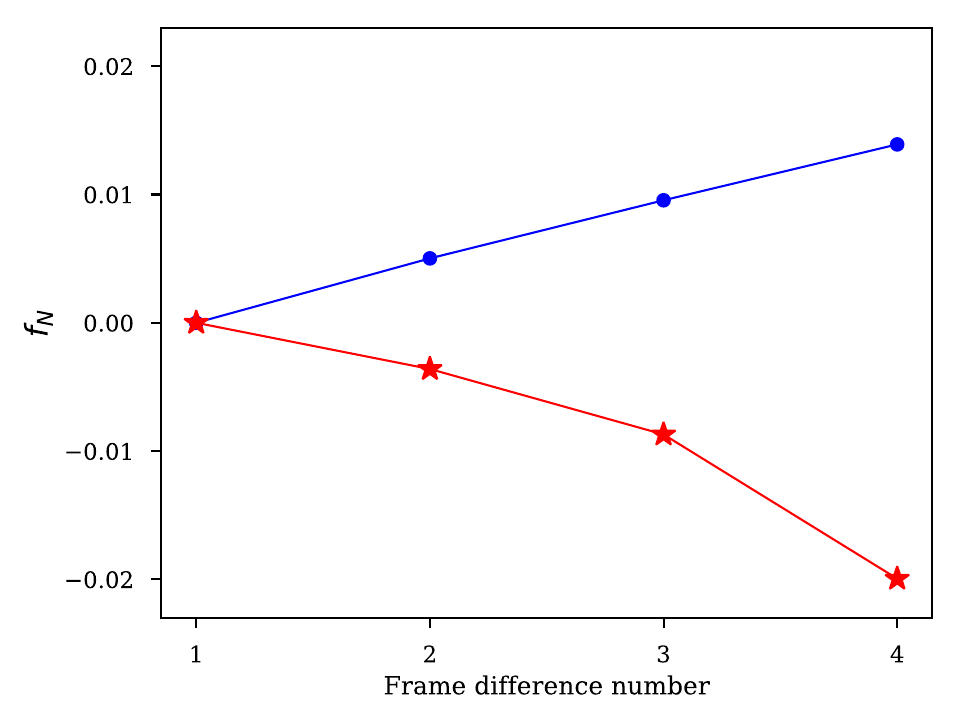}
\caption{\label{fig:charge} The upper line (blue, circles) represents the sum of the $f_N$ signal in each of the outer 8 pixels in Fig. \ref{fig:9pixels}. The lower line (red, stars) is the $f_N$ signal in the central pixel of Fig. \ref{fig:9pixels}. The similar amplitude (with opposite sign) indicates that charge is for the most part conserved (just redistributed) withing the 3$\times$3 postage stamp. The scale of the y-axis is the same as that of the central pixel in Fig. \ref{fig:9pixels}.}
\end{figure}

\begin{figure}[htbp]
\centering 
\includegraphics[width=1.0\textwidth, page=1]{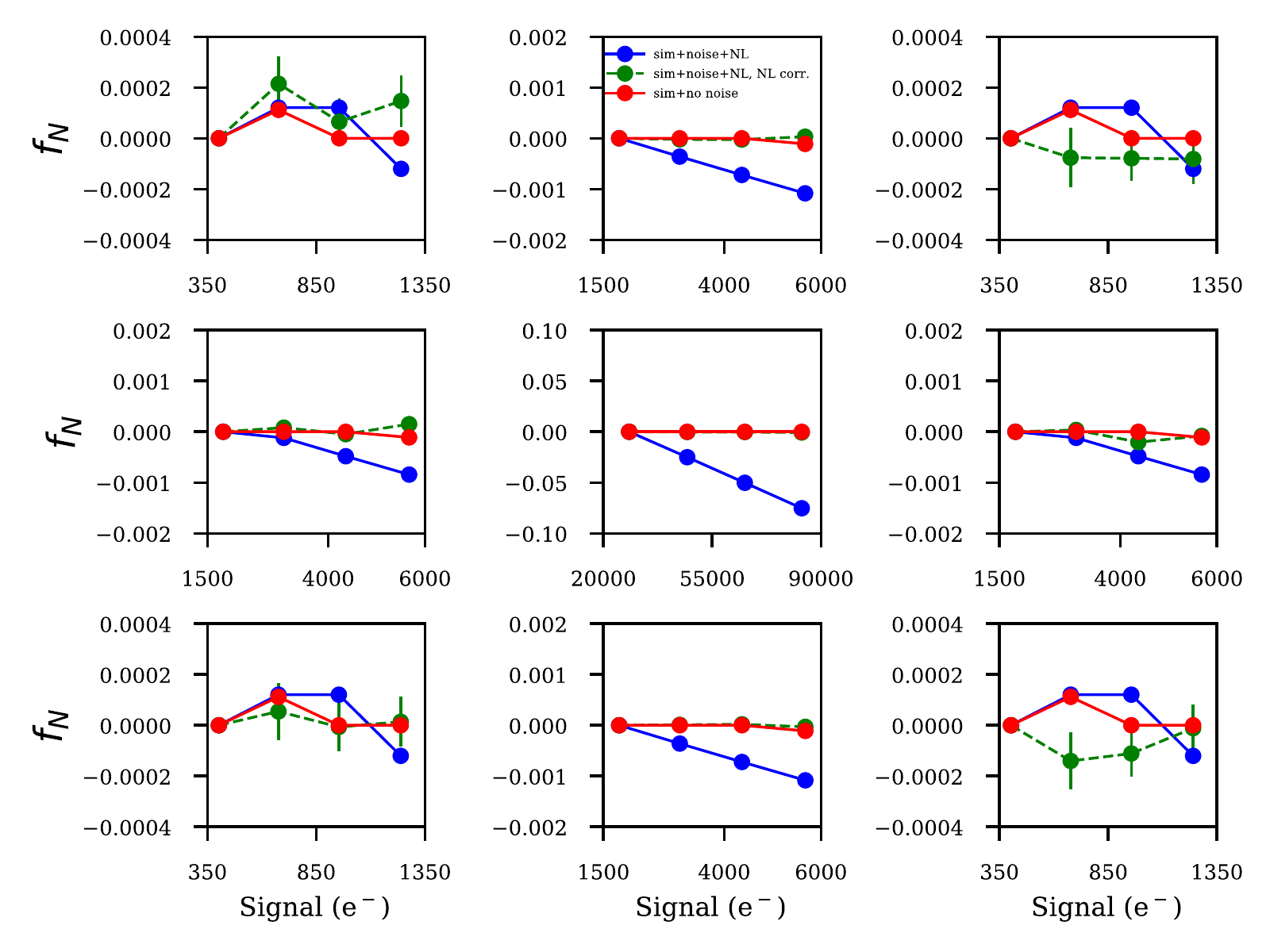}
\caption{\label{fig:9pixels_sim} Same as Fig. \ref{fig:9pixels}, but using simulations of centered spots (841 out of 1024 initial spots) instead of PPL data. Blue solid line: simulations with shot and read noise and classical NL applied (but not corrected), where $C_2$ follows a distribution of the form $\mathcal{N} \sim$ ($-7.76\times10^{-6}$, $0.36\times10^{-7}$). Green dashed line: same as blue, but with NL corrected with Eqn. \ref{eq:correction} as described in the text.}
\end{figure}

We quantify post-calibration nonlinearity in the stacked spot image by measuring  $f_N(k)$ for each pixel in the 3$\times$3 postage stamp around each spot, for each frame $k$.  We select spots with centroids within $0.1$ pixels of a pixel center to maximize the charge contrast between the central pixel and its nearest neighbors. We then calculate the average over the spots of the signal $f_N(k)$, after rejecting outliers through $3\sigma$ clipping. The final number of spots that satisfy these conditions is 748, and the results are shown in Fig. \ref{fig:9pixels} as a function of the average (over the 748 spots) of the mean integrated signal between the consecutive frames that define $F_k$ (the mean fluence of the $(k+1)$th and $k$th frames).  
As the exposures integrate, $f_N$ in the central pixel decreases, meaning its flux is lower relative to the beginning of the exposures by up to 2.1\% of the total spot flux.  This is accompanied by a simultaneous increase in the four nearest neighboring pixels, as would be expected from the shifting of effective pixel boundaries induced by a BF-type effect. As shown in Fig. \ref{fig:charge}, the excess flux in the eight outer pixels of the stamp is comparable to the flux loss in the central pixel; therefore, charge conservation in this region is nearly but not perfectly satisfied.
{As a check on our analysis pipeline}, Fig. \ref{fig:9pixels_sim} shows the $f_N$ values obtained when using simulated data (90 spots and flat ramps, each one with an initial grid of 1024 centered spots; 841 points remaining after analyzing the images with our measurement pipeline, which included rejection of sections of the detector where bad pixels lie in the actual lab data). The baseline case with noiseless images (and no IPC, NL, or BF effect) is consistent with a null result (red, solid line in Fig. \ref{fig:9pixels_sim}).
We then add noise and NL to the simulations, with NL applied in the form of a polynomial quadratic in the charge: 
\begin{equation}
Q (e)\rightarrow Q-\beta Q^2,
\end{equation}
with a second-order coefficient amplitude $\beta$ drawn from a Normal distribution of the form $\mathcal{N} \sim(-7.76\times 10^{-7},\ 0.36\times 10^{-7})$, consistent with measurements of NL on flat-field images at the laboratory. {IPC is also applied (as given by Eqn.\ref{eqn:k}), and subsequently deconvolved by the analysis pipeline, as would be done with actual data.}. If the effect of NL is not corrected by using Eqn. \ref{eq:correction}, the $f_N$ metric decreases as a function of time, with its amplitude being larger for the central pixel that concentrates more charge (Fig. \ref{fig:9pixels_sim}, blue solid line). When the correction is applied, the results are consistent with a null value for $f_N$ (Fig. \ref{fig:9pixels_sim}, green dashed line). 

Notice that the effect of NL on $f_N$ has the same, negative sign as integration time increases in each pixel (i.e., there is a decrease in the charge recorded, as expected), whereas in the case of the results with the PPL spots, the sign of the effect is opposite between the central pixel and its four nearest neighbors, as expected from charge redistribution caused by a BF-type effect. 
\begin{figure}[htbp]
\centering 
\includegraphics[width=1.0\textwidth, page=6]{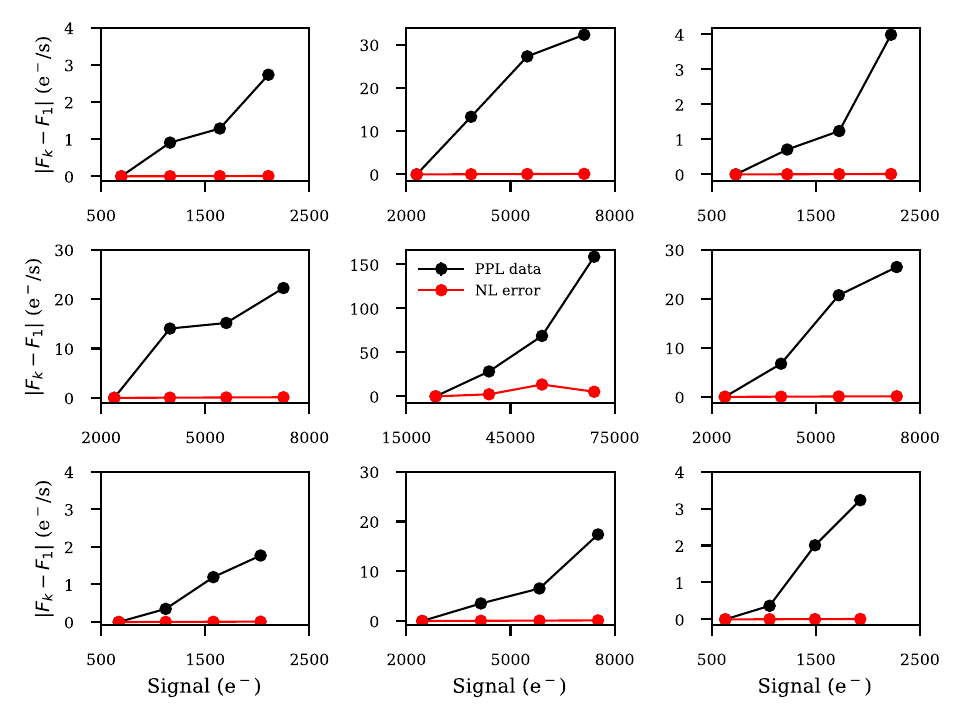}
\caption{\label{fig:9pixels_error} {Absolute value of flux changes per pixel relative to the beginning of a spot exposure.  Black points use the same spot data as in Fig. \ref{fig:9pixels} while red points show the estimated error due to the residuals of the NL model.  Flux $F_k$ is estimated from the differences between frames $k+1$ and $k$, and the x-axis is given by the average fluence between those frames.  We do not use the reset frame ($k=0$).  The NL error was estimated by modulating a simulated signal of the average spot in Fig: \ref{fig:spots} with the NL residual functions shown in Fig. \ref{fig:res}.}}
\end{figure}

{Since we are inferring the BF effect from changes in flux during an exposure, we must demonstrate that the residuals of our quadratic NL model are not the source of the observed flux changes.  To assess the impact of the residuals, we first simulate a median spot using the fluxes shown in Fig. \ref{fig:spots}. We then modulate each ramp from each pixel of the average spot with the NL residual functions from Fig. \ref{fig:res} to obtain the absolute fluence error expected from the NL residuals.  The residuals are linearly interpolated and extrapolated to low fluences by defining them to be zero at zero fluence.  We compute the change in flux induced between frames $F_k - F_1$ (i.e., the numerator of Eqn. \ref{eq:fn}) by the fluence errors, and compare them to the corresponding quantities measured from the PPL data (c.f., Fig. \ref{fig:9pixels} ). The results in Fig: \ref{fig:9pixels_error} show  the detected flux changes, which range from a few e$^-$/s in the corner pixels to 160 e$^-$/s in the center, are significantly larger than the expected errors due to insufficient calibration of NL.}
\begin{figure}[htbp]
\centering 
\includegraphics[width=1.0\textwidth, page=8]{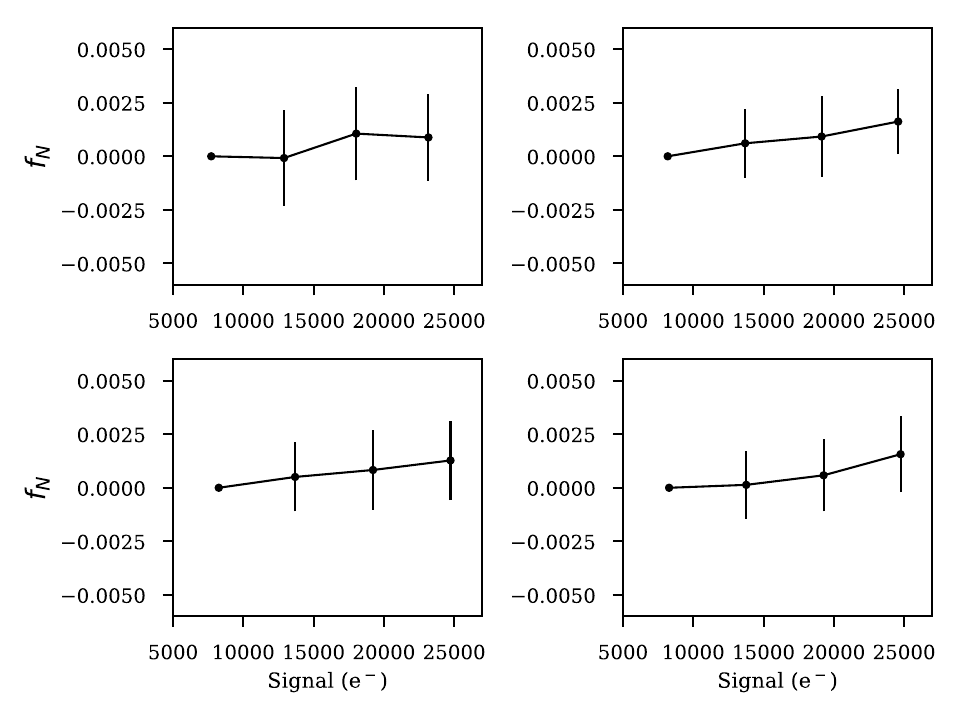}
\caption{\label{fig:corner} $f_N$ as a function of mean frame signal for PPL spots whose centroid lies within 0.1 pixels of the pixel corner (as opposed to the pixel center, c.f. Fig. \ref{fig:9pixels}). The charge is approximately evenly distributed among 4 pixels because the centroid is close to one of the four corners of the central pixel in the 3$\times$3 pixels region. For each case, we isolated the 4 pixels that contain the majority of the signal, and produced a stacked imaged by taking the mean of each of these four 2$\times$2 pixels subregions (i.e., translating them onto each other). The $f_N$ metric from this stacked image is shown here.The number of spots for each of the four cases is 40, 50, 43, and 50, respectively. Notice that the scale of the y-axis is the same as that of the 4 nearest neighbors of the central pixel in Fig. \ref{fig:9pixels}.}
\end{figure}

As a check on whether the BF effect depends on the signal contrast between adjacent pixels, we repeat our analysis on spots whose centroids are within 0.1 pixels of a pixel corner (with the charge approximately evenly distributed among four pixels), expecting an attenuation of the signal. Fig. \ref{fig:corner} shows the $f_N$ metric results for this particular case. As expected, the fluxes in each pixel are more constant (no change detected) during the exposures despite the spots having similar total fluxes to spots centered on a pixel.
\begin{figure}[htbp]
\centering 
\includegraphics[width=1.0\textwidth, page=3]{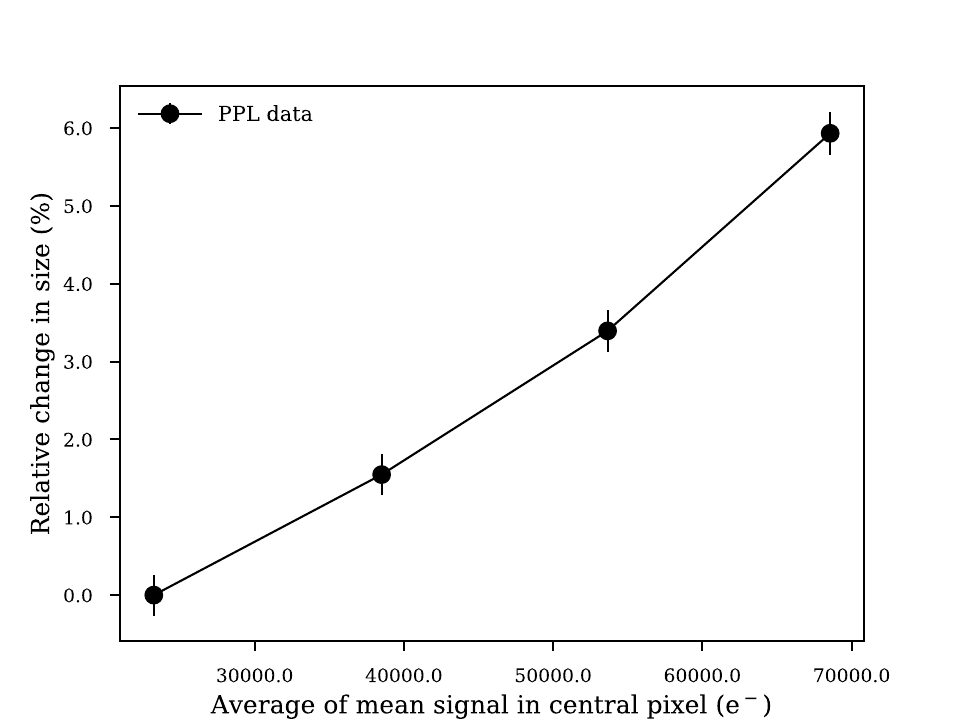}
\caption{\label{fig:size} Average relative change in size (second moment of the 3$\times$3 postage stamp) of the projected spots as a function of the average (over all the spots) of the mean signal in the central pixel between consecutive frames. The size is measured in the postage stamp formed by the difference of two consecutive frames, with the first difference being used as reference. The error bars represent the standard error of the mean value.}
\end{figure}

We also calculate the increase in the average relative size of the spots as the ramp is sampled in time. For the stamp formed as the difference between consecutive frames of a given ramp, we calculate the size $r$, defined as a linear combination of the second moments of the postage stamp $I$ weighted by a Gaussian of $\sigma = 1$ pixel: 
\begin{align}
r&=M_{11} + M_{22}  \\
M_{k,l}&=\frac{\sum_{i,j} x_{k,l} I_{i,j} w_{i,j}}{\sum_{i,j} I_{i,j} w_{i,j}} \nonumber
\label{eq:size}
\end{align}
Fig. \ref{fig:size} shows the normalized size (average over spots) with respect to the first frame difference (i.e., $r_k/r_1 -1$) as a function of {the average (over all the spots) of the mean signal in the central pixel between consecutive frames}, for the same PPL spots as in Fig. \ref{fig:9pixels}.  We detect a linear relative increase in the size of approximately $6\%$ as integration time increases ({c.f. with the 0.1\% or 10$^{-3}$ maximum change in relative change in size demanded by WL science in WFIRST}).

\subsection{{Estimating change in pixel area}}
{Assuming that the measured postcalibration nonlinearity is entirely due to an effective change in size of the central pixel in a spot  (ignoring other possible NL contributions such as nonlinear IPC), and assuming that the change in pixel area is proportional to charge contrast with nearest neighbor pixels, we can model the effect on flux in the central bright pixel as:  }
\begin{equation}
\frac{dQ}{dt} = F(t)=F_0 \left( 1 + B Q_{\mathrm{c}}(t) \right) =  F_0 \left( 1 + B F_{\mathrm{c}}t \right), 
\label{eq:f}
\end{equation}
where $F_0$ represents the unmodified flux in the pixel (in e$^{-}/s$), and $Q_{\mathrm{c}}$ and $F_{\mathrm{c}}$ are the mean charge and flux contrast between the central pixel and its four nearest neighbors, respectively: 
{\begin{align}
Q_{\mathrm{c}} &= {Q_{\mathrm{central}}} - { (Q_{\mathrm{left}} + Q_{\mathrm{right}} + Q_{\mathrm{top}} + Q_{\mathrm{bottom}} )/4 } \\
Q_{\mathrm{c}} &= F_{\mathrm{c}} t  \nonumber
\end{align}}
{For simplicity, we model the change in flux as being proportional to $F_0$, but more accurately, it would depend on the value of the spot profiles at the pixel boundaries being shifted.}  The  mean $F_{\mathrm{c}}$ is approximately constant, assuming that it is large compared to the amplitude of the BF effect. The proportionality constant $B$ quantifies the area change ($dA/A$) per e$^{-}$ of contrast $Q_{\mathrm{c}}$ (thus, the units of $B$ are 1/e$^{-}$). 
We calculate B by fitting the measured $f_N$ in the central pixel of each spot to a linear function of the form $f_N=mk$ (for frame number $k$ being). Setting $F_0$ to the flux in the first frame ($F_1$) and $t \equiv k \Delta t $ in Eqn. \ref{eq:f}, $f_N$ (Eqn. \ref{eq:fn}) can be rewritten as: 
\begin{equation}
f_N= \frac{B F_1 F_c \Delta t}{F_{*}}k
\end{equation}
Thus, the constant $B$ can be calculated using the slope from the linear fit as:
\begin{equation}
B = \frac{m}{F_{\mathrm{c}}} \left(\frac{F_{*}}{F_1 \Delta t} \right)
\label{eq:b}
\end{equation}

\begin{figure}[htbp]
\centering 
\includegraphics[width=1.0\textwidth, page=4]{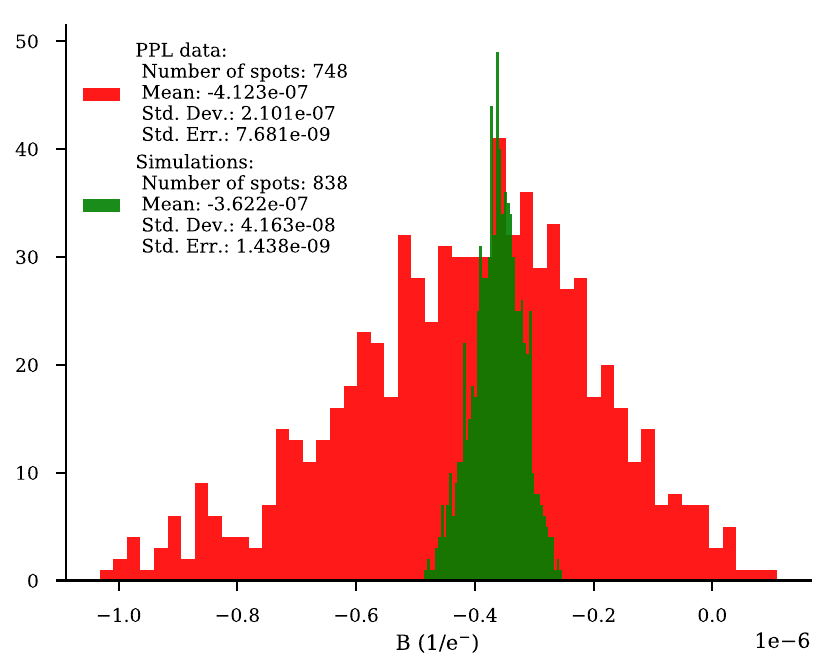}
\caption{Histogram of B coefficient as defined in Eqn. \ref{eq:b}: PPL data (red) and simulations with noise, NL, and BF effect applied (green).} 
\label{fig:hist_b}
\end{figure}

Fig. \ref{fig:hist_b} shows the distribution of $B$ coefficients as calculated for 748 spots that satisfied the centroid condition ($<$0.1 pixels from a pixel center). {The scatter of the distribution has contributions from random noise (dominated by shot noise), systematic errors, and the intrinsic scatter of the $B$ coefficient itself.} {To assess the contribution from shot noise}, we have simulated 90 ramps with shot noise and perfectly centered spots on a 2k by 2k grid, using the \citet{antilogus14} model in {\tt{GalSim}} to simulate the BF effect {corresponding to a single value of $B$}.\footnote {The parameters of the {\tt{GalSim}} model for the BF effect were chosen to resemble the measured mean:  {\tt{galsim.cdmodel.PowerLawCD (1, 1.1e-7, 1.1e-7, 1.0e-7, 1.0e-7, 0.0, 0.0, 0.0)}}. The first parameter indicates the maximum pixel separation out to which charges contribute to deflection, which we have set to 1 in this case.} The exact meaning of each parameter can be found in the {\tt{Galsim}} documentation. The resulting distribution of B from the simulations is shown in Fig. \ref{fig:hist_b} as well (green), and it shows that the scatter in the measured $B$ parameter (red) is not entirely due to noise but also has contribution from intrinsic scatter in $B$ or systematic errors, including possible errors in our model. The measured mean of the $B$ distribution is -0.41$\pm$0.0076 ppm/e$^-$ (the negative sign indicating shrinkage of the central pixel).  This provides an initial order of magnitude estimate of the BF effect, which can be refined by more in-depth treatment of systematics, and which may of course depend on the particular device and its operating conditions.



\section{Conclusions}

{We have directly demonstrated the existence of a BF effect in an H2RG-18 NIR detector by studying exposures of compact point sources generated at JPL's Precision Projector Laboratory. Using a stable and controlled laboratory environment, we extend and confirm the analysis in \citet{plazas17} (initiated by J. Anderson, Space Telescope Science Institute), which found preliminary evidence for a BF-type effect in Hubble/WFC3-IR data. 
We find evidence of charge redistribution from the central pixel of a bright point source to neighboring pixels after correcting for known effects such as classical voltage nonlinearity and linear IPC (assumed to be spatially uniform). Using 1 $\mu$m illumination and spots with FWHM=0.78 pixels, we measure a relative increase in the average spot size of about $5\%$ once the central bright pixel reaches 76000 e$^-$ (57\% of the mean pixel well depth). Our results are consistent with the physical concept proposed in \citet{plazas17}, that the shrinking of the depletion region in each pixel diode causes a contrast-dependent shift in the effective pixel boundaries. Assuming this mechanism, our measurements can be interpreted as a shrinking of the central bright pixel of the spot image, approximately parameterized as a 0.41 $\pm$ 0.0076 ppm reduction in area per electron of charge contrast with neighbor pixels. }

{If not corrected, the BF effect induces PSF shape measurement errors that will significantly degrade galaxy shape measurements in WL science analyses using HxRG detectors, such as WFIRST which requires PSF sizes measured to 0.1\%.  Shifting pixel boundaries may also cause photometric and astrometric errors in NIR supernova and microlensing measurements, especially in crowded fields where measurements may be influenced by nearby bright sources.
Future work should further characterize and model the effect, exploring dependence on parameters such as illumination wavelength, integration time, and source size/shape.  Sensitivity to detector bias voltages should also be studied to understand the physics of the effect.  Of course we expect the BF effect to be device-dependent: the results here are not fully applicable to the H4RG-10 detectors (10 $\mu$m pitch) planned for the WFIRST Wide Field Instrument, but they are likely to be qualitatively similar.  It will be crucial for WFIRST and other astronomy missions using HxRG detectors (e.g. James Webb Space Telescope and the Thirty Meter Telescope) to not only characterize the BF effect but also to validate that precision measurements (shapes, astrometry, and photometry) can be recovered after applying an intended correction scheme on representative devices.  The PPL detector emulation facility is well-suited for studies of this type thanks to the speed, stability, and versatility of the projector testbed.
}


\section*{Acknowledgements}

We thank Chris Hirata, Jeff Kruk, Dave Content, Mike Seiffert, Warren Holmes, the WFIRST detector requirements working group, the Euclid detector working group, and Goddard's Detector Characterization Laboratory group for useful discussions. We thank Stefanie Wachter for useful discussions and feedback on the initial drafts of this document. We thank Warren Holmes and the Euclid detector working group for providing the H2RG detector used in this work. We acknowledge funding from the WFIRST and Euclid projects. AAP is supported by the Jet Propulsion Laboratory. AAP acknowledges support from the Italian Ministry of Foreign Affairs and International Cooperation, Directorate General for Country Promotion. CS, JR, and EH are being supported in part by the Jet Propulsion Laboratory. The research was carried out at the Jet Propulsion Laboratory, California Institute of Technology, under a contract with the National Aeronautics and Space Administration. 

\textcopyright 2017. All rights reserved.
\bibliographystyle{plainnat}
\bibliography{bf_bib}

\end{document}